\documentclass[twocolumn,pre,showpacs,amsmath,amssymb,floatfix]{revtex4-1}

\usepackage[colorlinks=true,citecolor=blue,linkcolor=blue]{hyperref}
\usepackage{natbib}
\usepackage{amssymb}
\usepackage{amsmath}
\usepackage{graphicx}

\newcommand{\parder}[3]{\left(\displaystyle\frac{\partial #1}{\partial #2} \right)_{#3}}

\newcommand{\nn}{{\tilde n}}
\newcommand{\ww}{{\tilde w}}

\newcommand{\ens}[1]{\left\langle #1 \right\rangle}
\newcommand{\sol}{\text{sol}}
\newcommand{\gel}{\text{gel}}
\newcommand{\kk}[1]{{\bar k_\mathbf{#1}}}
\newcommand{\kens}[1]{\ens{k_{#1}}}

\begin{document}
\title{Statistical Thermodynamics of Irreversible Aggregation and Gelation}%
\author{Themis Matsoukas}
\email{matsoukas@psu.edu}
\affiliation{%
Department of Chemical Engineering, Pennsylvania State University, University Park, PA 16802}%
\date{\today}

\begin{abstract} 
Binary aggregation is known to lead, under certain kinetic rules, to the coexistence of two populations, one consisting of finite-size clusters (sol), and one that contains a single cluster that carries a finite fraction of the total mass (giant component or gel). The sol-gel transition is commonly discussed as a phase transition by qualitative analogy to vapor condensation. Here we show that the connection to thermodynamic phase transition is rigorous. We develop the statistical thermodynamics of irreversible binary aggregation in discrete finite systems, obtain the partition function for arbitrary kernel, and show that the emergence of the gel cluster has all the hallmarks of a phase transition, including an unstable van der Waals loop. We demonstrate the theory by presenting the complete pre- and post-gel solution for aggregation with the product kernel, $k_{ij}=i j$.

\end{abstract}
\maketitle

\section{INTRODUCTION}
A ubiquitous problem in the physics of dispersed systems is binary aggregation: two clusters $i$ and $j$ merge with probability proportional to the aggregation kernel $k_{ij}$, a function that characterizes the physics of $i$-$j$ encounters. This process describes many physical phenomena over a length scales that encompass molecular systems, social networks, and stars \cite{Leyvraz:PR03,Stockmayer:JCP43}. 
Under certain kinetic rate laws that preferentially promote the merging of large clusters, this process produces a remarkable behavior, a \textit{phase transition} manifested in the emergence of a single element that contains a finite fraction of the members of the population.  This transition is seen in experimental systems (gelation) as well in dynamic stochastic models, most notably percolation \cite{Aldous:B99}. 
The standard mathematical tool is Smoluchowski's coagulation equation, developed nearly 100 years ago \cite{Smoluchowski:ZFPC17,Smoluchowski:PZ16}. It is an integral-differential equation that governs the evolution of the mean cluster size distribution of an infinite system whose total mass is fixed. The Smoluchowski equation forms the basis for the quantitative study of colloids and polymers, atmospheric aerosols, animal populations, and dispersed populations in general \cite{Stockmayer:JCP43,Friedlander:00,Gueron:MB95} and its mathematical behavior has been studied extensively  \citep{Drake:72,Aldous:B99,Leyvraz:PR03}. 

The product kernel $k_{ij}=ij$ is a classical example of a kernel that produces a giant cluster within finite aggregation time \cite{Leyvraz:PR03}. It is a model for polymer gelation (polymerization of $f$-functional monomers in the limit $f\to\infty$ \citep{Flory:JACS41b,Stockmayer:JCP43}) and for percolation on random graphs \citep{Erdos:PMIHA60,KrapivskyRednerBenNaim}. As such, it serves as the standard analytic model for the study of the giant component. 
In the presence of the giant cluster (``gel phase'') the Smoluchowski equation breaks down: the second moment of the size distribution diverges at the gel point (the divergence defines the gel point), and past this point the first moment decays, i.e., mass is not conserved.  To restore consistency one assumes the presence of a gel phase (not predicted by the Smoluchowski equation itself) and introduces an additional assumption as to whether the finite-size clusters (``sol phase'') interact with the gel (Flory model) or not (Stockmayer model) \citep{Ziff:JCP80}. These heuristic assumptions lead to different solutions each. 
%

\begin{figure*}
\begin{center}
\includegraphics[width=6.25in]{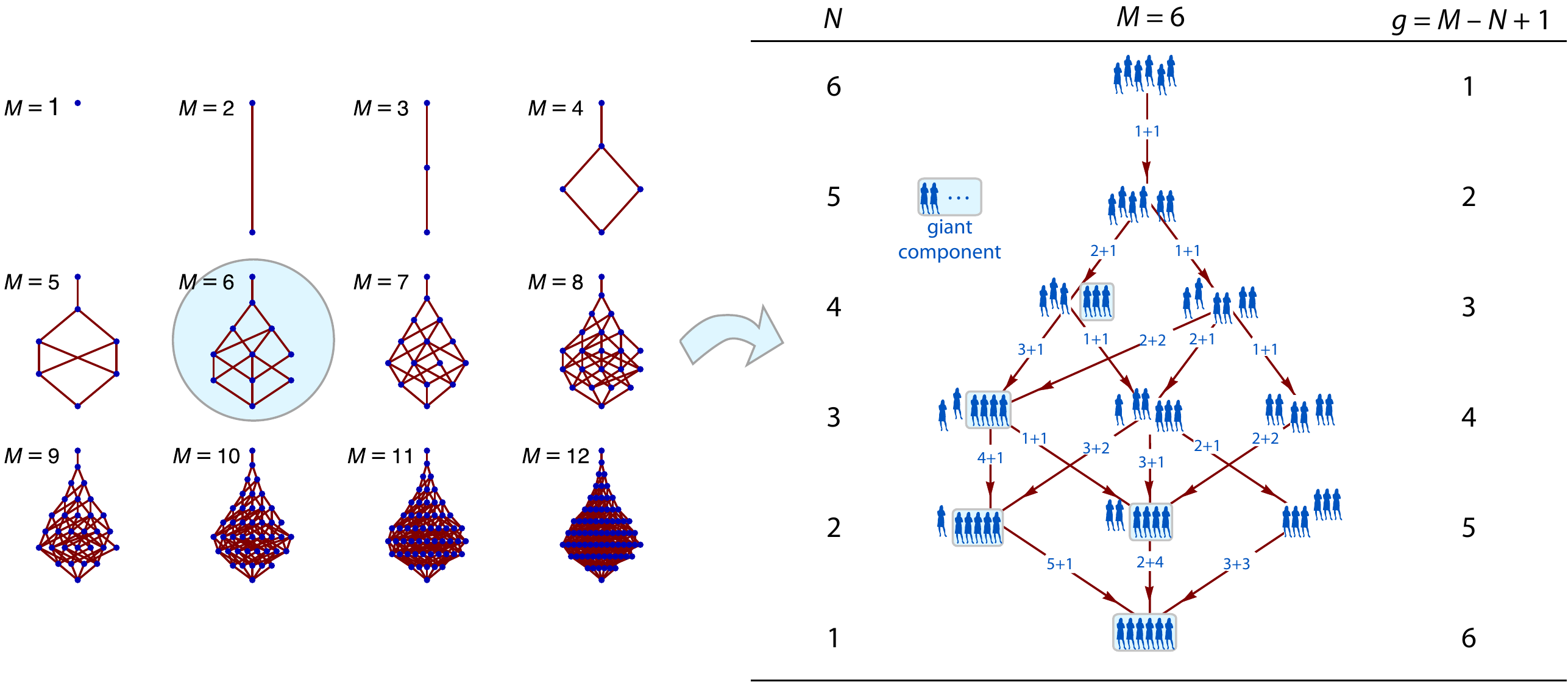}
\end{center}
\caption{The ensemble of discrete binary aggregation graphs ($M=1$ through 12 shown). Each graph starts with $M$ unattached monomers at the top and ends with a single cluster at the bottom. The graph $M=6$ is shown in detail on the right. Nodes represent distributions; arrows point from the parent distribution to the offspring and are labelled by the sizes of the two clusters whose merging produces the offspring. All distributions at fixed $N$  constitute the $(M,N)$ ensemble.  The giant component (highlighted) is identified as the cluster in the size region $i>(M-N+1)/2+1$. At most one cluster can exist in the giant region \citep{Matsoukas:statPBE_PRE14}. }
\label{fig1}
\end{figure*}

The limitation of the Smoluchowski equation arises from the fact that it reduces an inherently discrete finite stochastic process into a single metric, the mean cluster size distribution. The need for stochastic treatments has been raised in the literature. The discrete finite problem was originally formulated by Marcus \cite{Marcus:T68} and studied in detail by Lushnikov, who obtained the discrete size spectrum for several special kernels, including the complete self-consistent solution (pre- and post gel) for the product kernel $k_{ij}=i j$.  \cite{Lushnikov:JCIS78,Lushnikov:PRL04,Lushnikov:PRE05}.  Spouge \cite{Spouge:M83_831} and Hendriks et al.~\cite{Hendriks:ZPCM85B} used a combinatorial approach to obtain solutions for special, non-gelling cases. Other discrete, branching solutions have been implemented in closely related problems in aggregation and fragmentation \cite{Durrett:JTP99,Berestycki:07}, stochastic structures \cite{Aldous:B99}, biosciences \cite{Duerr:MB00} and astrophysics \cite{Sheth:MNRAS97}. 
These treatments gives results for several special cases, but none offers the physical context in which to address the main question: what is the nature of sol-gel transition and how it relates to familiar phase transitions?  We are now in position to make this  connection. Recently we formulated the statistical thermodynamics of generic populations \citep{Matsoukas:statPBE_PRE14}. Here we apply this theory to irreversible aggregation, develop the thermodynamics the discrete finite domain, and obtain the solution to the product kernel.

\section{Aggregation in Discrete Finite Systems}
We cast the problem in the theory of the cluster ensemble \citep{Matsoukas:statPBE_PRE14}. 
We consider a population of $M$ individuals (``monomers'') that form $N$ clusters and construct the microcanonical ensemble of all possible distributions $\mathbf{n}=(n_1,n_2,\cdots)$, where $n_i$ is the number of clusters with $i$ monomers. All distributions of the $(M,N)$ ensemble satisfy the two constraints
\begin{equation}\label{constraints}
   \sum n_i = N,\quad \sum i n_i = M . 
\end{equation}
When two clusters in distribution $\mathbf{n}$ of the $(M,N)$ ensemble merge, the outcome is a new distribution in the ensemble $(M,N-1)$ of the next generation. We formally define generation $g=M-N+1$ such that $g=1$ refers to fully dispersed monomers and $g=M$ to a fully gelled state. These parent-offspring relationships  produce a directed graph that represents the phase space of discrete binary aggregation (Fig.\ \ref{fig1}).  Following \citep{Matsoukas:statPBE_PRE14}, we express the probability of distribution $\mathbf{n}$ in the $(M,N)$ ensemble as  
\begin{equation}\label{Prob}
   P(\mathbf{n}) = \mathbf{n!}W(\mathbf{n})/\Omega_{M,N},
\end{equation}
where $\mathbf{n!}=N!/n_1!\, n_2!\,\cdots$ is the multinomial coefficient,  $\Omega_{M,N}$ is the partition function, and $W(\mathbf{n})$ is the bias of distribution, a functional of $\mathbf{n}$ that is determined by the physics of the problem, here, by $k_{ij}$. As shown in \citep{Matsoukas:statPBE_PRE14}, the most probable distribution (MPD) in the thermodynamic maximizes $\Omega_{M,N}$, and in the thermodynamic limit is given by 
\begin{equation}\label{mpd}
   \frac{\nn_i}{N} = \ww_i \frac{e^{-\beta i}}{q}
\end{equation}
where $\beta$, $\log q$ and $\log\ww_i$ are given by the partial derivatives,
\begin{align}
\label{beta}
   & \beta     =\parder{\log\Omega_{M,N}}{M}{N},\\
\label{logq}
   & \log q    =\parder{\log\Omega_{M,N}}{N}{M},\\
\label{wwi}
   & \log\ww_i =\parder{\log W(\mathbf{\nn})}{\nn_i}{\nn_j}. 
\end{align}
We now seek the evolution of $\Omega_{M,N}$ and $W(\mathbf{n})$.   We begin with the probability $P(\mathbf{n})$, which propagates from parents to offsprings according to the time-free Master Equation 
\begin{equation}\label{agg_MasterEquation}
    P(\mathbf{n}) = 
          \sum_{\mathbf{n'}\rightarrow\textbf{n}} P(\mathbf{n'})
          P_{\mathbf{n'}\to\mathbf{n}}  ,
\end{equation}
where $\mathbf{n'}$ is the parent  that produces $\mathbf{n}$ by merging cluster sizes $i-j$ and $j$. The transition probability $P_{\mathbf{n'}\to\mathbf{n}}$ is given by
\begin{equation}\label{agg_P:transition}
   P_{\mathbf{n'}\to\mathbf{n}} = \frac{2}{N (N+1) \kk{n'}}
   \frac{n'_{i-j} (n'_j - \delta_{i-j,j}) }  {1+\delta_{i-j,j}} \, k_{i-j,j} , 
\end{equation}
where $\kk{n'}$ is the mean aggregation kernel among all pairs of clusters in parent $\mathbf{n'}$. Combining (\ref{Prob}), (\ref{agg_P:transition}) and (\ref{agg_MasterEquation}), we obtain the recursion
\begin{multline} 
\label{recursion_PFW}
   \frac{\Omega_{M,N+1}}{\Omega_{M,N}} = 
   \frac{M-N}{N}\frac{1}{\kens{M,N+1}}
\times\\
   \left\{
   \frac{\ens{k_{M,N+1}}}{M-N}
   \sum_{i=2}^\infty n_i\sum_{j=1}^{i-1}
   \frac{k_{i-j,j}}{\kk{n'}}\,\frac{W(\mathbf{n'})}{W(\mathbf{n})}
   \right\} , 
\end{multline}
where $\ens{k_{M,N+1}} = \sum P(\mathbf{n'})\kk{n'}$ is the ensemble average kernel over all distributions of the parent ensemble.  
Equation (\ref{recursion_PFW}) applies to all distributions $\mathbf{n}$ of the $(M,N)$ ensemble, and since the left-hand side is strictly a function of $M$ and $N$, the same must be true for the right-hand side:  the quantity in braces must be independent of $\mathbf{n}$. We further require homogeneous behavior in the thermodynamic limit, such that $\log\ww_i$ is an intensive function of $M/N$ \citep{Matsoukas:statPBE2}. This condition fixes the quantity in braces to be 1 and breaks Eq.\  (\ref{recursion_PFW}) into two separate recursions, one for $\Omega$ and one for $W(\mathbf{n})$. The first recursion  is 
\begin{equation}\label{recursion_PF}
   \frac{\Omega_{M,N+1}}{\Omega_{M,N}} = 
   \frac{M-N}{N}\frac{1}{\kens{M,N+1}}
\end{equation}
and is readily inverted to produce
\begin{equation}\label{PF}
   \Omega_{M,N} = \binom{M-1}{N-1}\prod_{L=N+1}^M\ens{k_{M,L}} .
\end{equation}
Thus we have the partition function in terms of $M$, $N$, and the product of all $\ens{k}$ from generation 1 up to the parent of the current generation.  We note that the binomial factor is the total number of ordered partitions of integer $M$ into $N$ \citep{Pitman:06}, also equal to the number of distributions in the $(M,N)$ ensemble, each counted $\mathbf{n!}$ times. The second recursion gives the selection bias $W(\mathbf{n})$ in terms of the bias of all parents of $\mathbf{n}$:
\begin{equation}\label{recursion_W}
   W(\mathbf{n}) =    \frac{\ens{k_{M,N+1}}}{M-N}
   \sum_{i=2}^\infty n_i\sum_{j=1}^{i-1}
   \frac{k_{i-j,j}}{\kk{n'}}\,W(\mathbf{n'}) . 
\end{equation}
Starting with $W=1$ in generation 1 we may obtain, in principle, the bias of any distribution in the phase space. Returning to Eq.\ (\ref{recursion_PF}), we recognize  the right-hand side as $q$, which produces the path equation of the process:
\begin{equation}\label{gov_q}
   q = \frac{M-N}{N}\frac{1}{\kens{M,N+1}} .
\end{equation}
Equations (\ref{PF}) and (\ref{recursion_W}), along with (\ref{mpd})--(\ref{wwi}) constitute a closed set of equations for the MPD in the thermodynamic limit.

\section{PRODUCT KERNEL}
We now apply the theory to obtain the solution to the product kernel $k_{ij}=i j$. In the thermodynamic limit, $\ens{k_{M,N}}\to\kk{n}\to (M/N)^2$. With this result and  Eqs.\ (\ref{beta})--(\ref{logq}) we obtain the parameters of the sol:
\begin{align}
\label{beta_prod}
   \beta &= 2\theta-\log \theta,\\
\label{q_prod}
   q     &= \theta(1-\theta),\\
\label{wwi_prod}
   \ww_i &=2 \frac{(2 i)^{i-2}}{i!} ,
\end{align}
with $\theta=1-N/M$.  The MPD follows from Eq.\ (\ref{mpd}):
\begin{equation}\label{mpd_prod}
     \frac{\nn_i}{N} = 
     \frac{2\theta}{1-\theta}
     \frac{(2\theta k)^{k-2}}{k!} e^{-2\theta k} . 
\end{equation}
The MPD of the sol is provided that $d^2\log\Omega_{M,N}\leq 0$, or $(\partial\log q/\partial N)_M\leq 0$ \citep{Matsoukas:statPBE2}.  Applying this stability condition to Eq.\ (\ref{q_prod}) we conclude that the range of stability is $0\leq\theta<1/2$ and that phase splitting must occur at $N^*=M/2$. This is the same as the gel point in the Smoluchowski equation with monodisperse conditions.  We now proceed to obtain solutions in the post-gel region.  
Consider a two-phase system that contains mass $M_\sol$ in the sol, and $M_\gel=M-M_\sol$ in the gel ($N_\gel=1$, $N_\sol=N-1$ \citep{Matsoukas:statPBE_PRE14}). As an equilibrium phase, the sol maximizes $\Omega_{M_\sol,N-1}\approx \Omega_{M_\sol,N}$. Its distribution, therefore, is given by Eq.\ (\ref{mpd}) with $\theta$ replaced by $\theta_\sol=1-N/M_\sol$. To determine $M_\sol$ we recall that Eq.\ (\ref{q_prod}) must be satisfied, at all times. Since stability requires $N\leq M/2$, we must have $M_\sol= M N/(M-N)$. Finally, the gel fraction is $\phi_\gel = (M-M_\sol)/M$, or
\begin{equation}\label{phi_gel}
   \phi_\gel = 2-1/\theta,\quad (\theta\geq 1/2). 
\end{equation}
Thus we have the complete solution: in the pre-gel region ($\theta\leq 1/2$) the sol is given by Eq.\ (\ref{mpd_prod}); in the post-gel region ($\theta\geq 1/2$) it given by the same equation with $\theta$ replaced by $\theta_\sol=1-\theta$, and the gel fraction is obtained from Eq.\ (\ref{phi_gel}). 
\begin{figure}[tb]
\begin{center}
\includegraphics[width=\columnwidth]{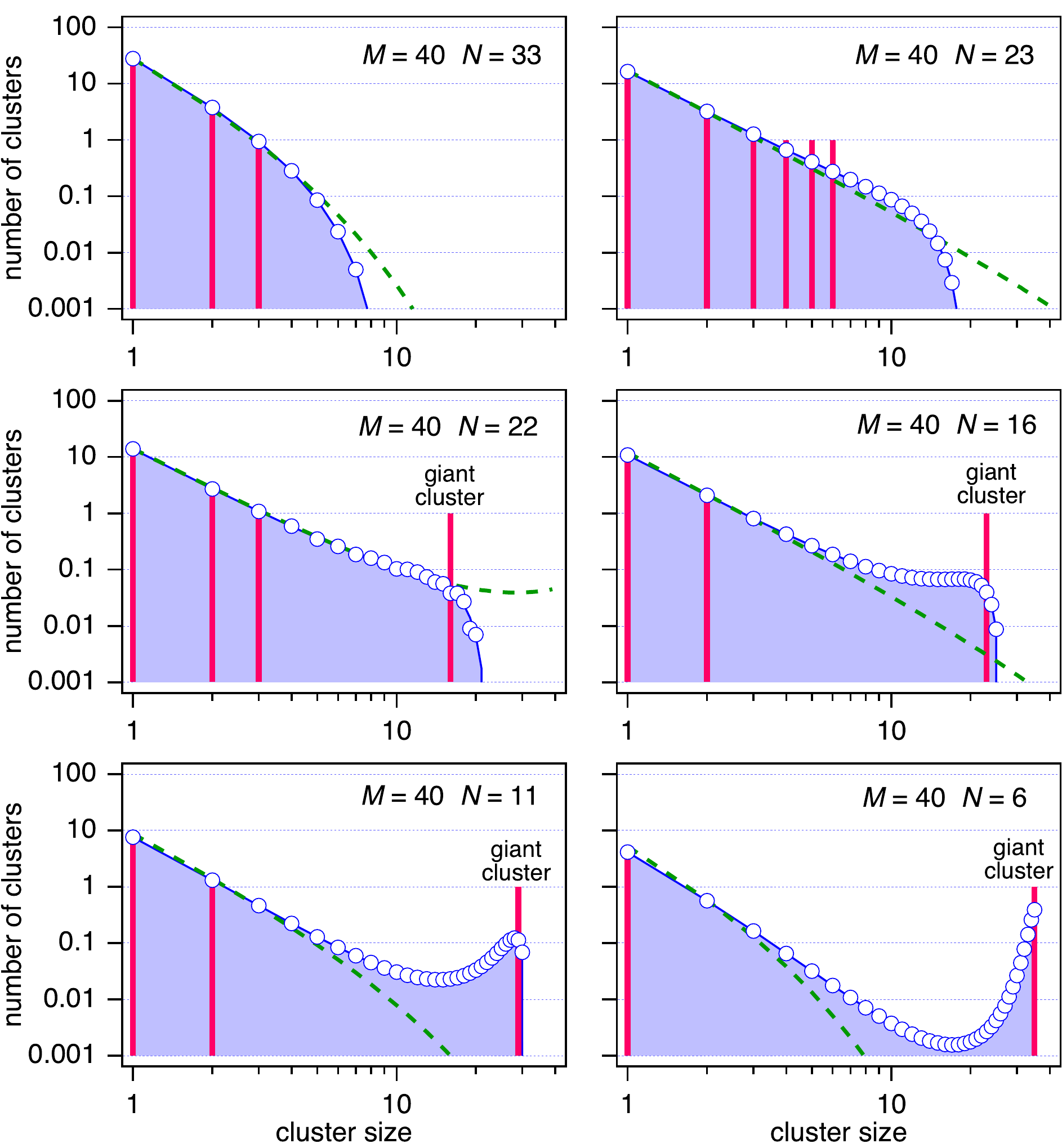}
\end{center}
\caption{Evolution of distributions in a population with $M=40$. Shaded curve: mean distribution (exact calculation by direct enumeration of all distributions); vertical sticks: most probable distribution (exact calculation); symbols: Monte Carlo simulation (average of 5000 repetitions); dashed line:  Eq.\ (\ref{mpd_prod}) (thermodynamic limit) with $\theta=1-N/M$ in the pre-gel region ($N\geq M/2$), and $\theta=N/M$ in the post-gel region.}
\label{fig2}
\end{figure}

%
We illustrate the theory with a numerical calculation for $M=40$. This value is sufficiently small that we may enumerate all  distributions on the aggregation graph and perform an exact calculation of the entire ensemble, yet large enough  that the thermodynamic limit is approached to satisfactory degree (the phase space contains 37338 distributions). As a further test we conduct Monte Carlo (MC) simulations by the constant-volume method \citep{Matsoukas:CES98}. The simulations sample the vicinity of the MPD (not the MPD itself) from which the mean distribution is calculated.  
The exact calculation is done on the entire graph as follows. Starting with $W=1$ in generation $g=1$, we apply Eq.\ (\ref{recursion_W}) to obtain the bias of all distributions in the next generation until the entire graph is computed. Next we calculate the partition function in each generation from the normalization condition $\Omega = \sum \mathbf{n!} W(\mathbf{n})$, and the probability of distribution from Eq.\ (\ref{Prob}). With all probabilities known, the mean distribution and the ensemble average kernel are readily calculated, and the MPD is identified by locating the maximum $P(\mathbf{n})$. As a check, we calculate the partition function from Eq.\ (\ref{PF}) and confirm that for pre-gel states it agrees with the result from the normalization condition. 

These calculations are compared in Figure \ref{fig2}, which shows selected distributions ranging from $N=33$ (early stage of mostly small clusters) to $N=6$  (nearly fully gelled). Since the MPD is an actual member of the ensemble, it contains integer numbers of clusters. The mean distribution is a composite of the entire ensemble and is not restricted to integer values. The giant cluster forms at $N^*=22$ and its presence is seen very clearly in the MPD. The gel phase is less prominent in the mean distribution because its peak is smeared by lateral fluctuations. Not all distributions in the vicinity of the MPD contain a giant cluster; as a result, the gel fraction grows smoothy at the gel point. 
In the sol region $1\leq (M-N+1)/2$, the theoretical distribution from Eq.\ (\ref{mpd_prod}) and the mean distribution are in excellent agreement.  The analytic result eventually breaks down when $N\to 1$ (the thermodynamic limit is violated at this point), yet even with $N$ as small as 6, agreement with theory remains acceptable.  The mean distribution from MC is practically indistinguishable from that by the exact calculation.  This confirms the validity of Eq.\ (\ref{recursion_W}), which forms the basis of the exact calculation. 

\begin{figure}[t]
\begin{center}
\includegraphics[width=0.9\columnwidth]{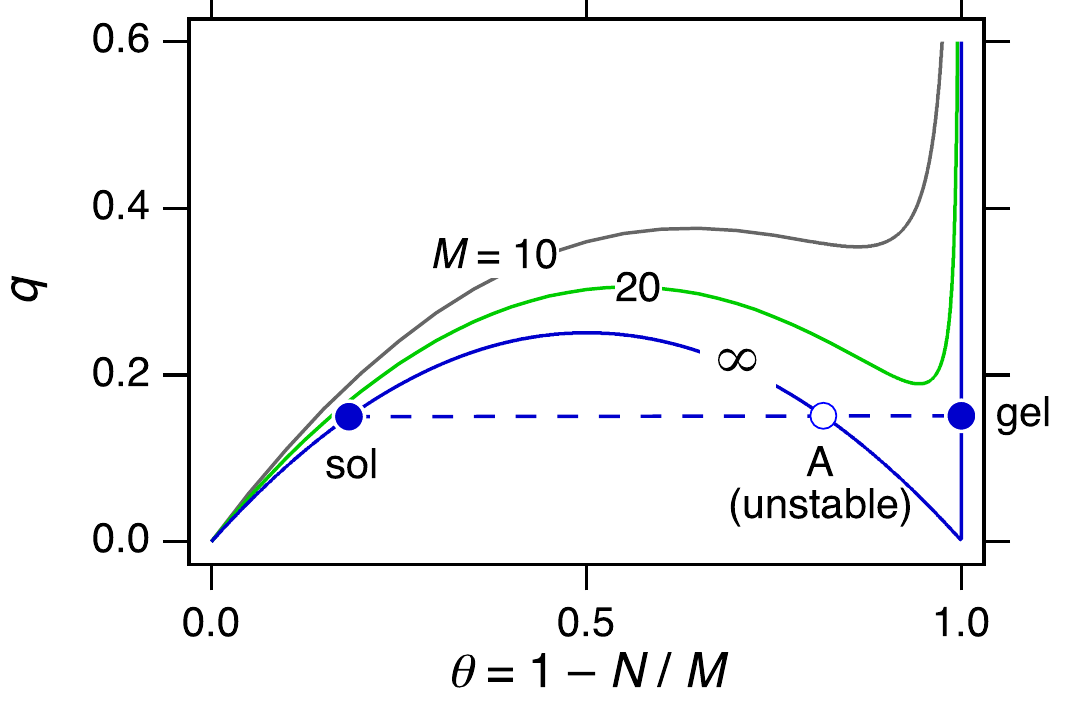}
\end{center}
\caption{Graphical constructions of tie line in the sol-gel region.  State $A$ is unstable and splits into a sol and a gel phase. All three states lie on the same tie line. }
\label{fig3}
\end{figure}
\section{DISCUSSION}
Our results make contact with several studies in the literature.  A recursion for the  partition function that is similar to that in Eq.\ (\ref{recursion_PF}) (different by a factor that is inconsequential for the statistics of the ensemble but crucial for thermodynamics to work) was obtained by Spouge \cite{Spouge:JPMG85,Spouge:M83_121,Spouge:M83_831} by a  combinatorial derivation for kernels of the form $k_{ij}= \alpha + \beta (i+j)+\gamma(i j)$ in pre-gel states. 
We recognize Eq.\ (\ref{mpd_prod}) as the classical pre-gel solution to the Smoluchowski equation \cite{Leyvraz:PR03,Aldous:B99}. We further recognize the post-gel solution as the Flory model, which assumes that the sol fraction continues to interact with the giant component past the gel point \cite{Ziff:JCP80}. No such \textit{a priori} assumption is required here: as long as no cluster is excluded from merging, a condition already built into the kernel ($k_{ij}\neq 0$ for all $i,j\geq 1$), the post-gel solution \textit{is} the Flory solution.  

We close with a final observation that points to an even closer analogy to molecular systems. Using Eq.\ (\ref{gov_q}) to calculate $q$ we find
\begin{equation}\label{q:loop}
   q = \left(\frac{M-N}{N}\right)\left(\frac{N+1}{M}\right)^2,
\end{equation}
whose limiting value for $M\gg 1$ is the result given in Eq.\ (\ref{q_prod}). Plotted against $\theta=1-N/M$ over the full range $\theta=0$ to 1, this equation shows behavior reminiscent of subcritical van der Waals isotherms (Fig.\ \ref{fig3}) and for large $M$ it converges to a parabola in the region $0\leq \theta < 1$, plus a Dirac delta function at $\theta=1$. Stability requires $(\partial q/\partial\theta)_M\geq 0$, a condition that is met in  $0\leq\theta\leq1/2$, but also on the Dirac branch. 
When the system crosses into the unstable region (state $A$ in Fig.\ \ref{fig3}) it must split into two phases. The sol phase is determined by Eq.\ (\ref{gov_q}), which produces a state on the stable branch ($\theta<1/2$) at the same $q$. Extending this line to the right we obtain an intersection with the Dirac delta branch, which we identify as the equilibrium gel phase at $\theta_\gel=1$. Thus we have the tie line of this two-phase system: it connects two equilibrium phases, with an unstable state at the middle.

\section{Conclusions}
The ensemble method was applied here to binary aggregation but can be adapted to any other growth mechanism. For example, by reversing the arrows in Fig.\ \ref{fig1} we obtain the graph of binary fragmentation; by including both directions we obtain the graph for reversible aggregation/fragmentation (both processes share the same trajectories in phase space as binary aggregation). In general, the evolution of populations may be viewed as a swarm of trajectories in the phase space of Fig.\ (\ref{fig1}) under parent-offspring relationships that must be derived for each case separately.  We may draw, therefore, a rigorous connection between statistical thermodynamics and population balances that offers new insights into the dynamics of evolving populations. 
\appendix
\section{Derivations}
\subsection{Derivation of Recursions, Eqs.\ (\ref{recursion_PF}) and (\ref{recursion_W})}
First write Eq.\ (\ref{recursion_PFW}) in the form
\begin{align}
\label{app_fund_r}
   &\frac{\Omega_{M,N+1}}{\Omega_{M,N}} 
      = \frac{M-N}{N}\frac{A}{\ens{k_{M,N+1}}}, \\
\label{app_A_def}
   &A = \frac{1}{M-N}\sum_{i=2}^\infty n_i \sum_{j=1}^{i-1}
        \frac{k_{i-j,j}}{\kk{n'}/\ens{k_{M,N+1}}}
        \frac{W(\mathbf{n'})}{W(\mathbf{n})} .
\end{align}
Since the LHS in Eq.\ (\ref{app_fund_r}) is independent of distribution $\mathbf{n}$, $A$ must be an (intensive) ensemble property. In the thermodynamic limit the bias of parents and offsprings satisfies the homogeneity condition \citep{Matsoukas:statPBE_PRE14}
\begin{equation}\label{app_homo}
   \log W(\mathbf{n}) = \sum_i n_i\log\ww_i 
\end{equation}
from which we obtain
\begin{equation}
   \frac{W(\mathbf{n'})}{W(\mathbf{n})} = \frac{\ww_{i-j}\ww_j}{\ww_i}. 
\end{equation}
We use the above result to express Eq.\ (\ref{app_A_def}) in the form
\begin{align}
\label{app_A_a}
   A &= \frac{1}{M-N}\sum_{i=2}^\infty n_i (i-1)\,a_i\\
\label{app_a_def}
   a_i &= \frac{1}{i-1}\sum_{j=1}^{i-1}
        \frac{k_{i-j,j}}{\kk{n'}/\ens{k_{M,N+1}}}
        \frac{\ww_{i-j}\ww_j}{\ww_i} .
\end{align}
Here $\mathbf{n'}$ is the parent distribution that produces the offspring distribution $\mathbf{n}$ by aggregation of the cluster pair $(i-j) + (j)$. In the thermodynamic limit ($\mathbf{n}\to\mathbf{\nn}$), the ratio $\kk{n'}/\ens{k_{M,N+1}}$ is independent of the individual parent and becomes a function of $i-j$ and $j$. The $j$ summation then produces a result, $a_i$, that is a function of $i$, $M$ and $N$.  The set $(a_1,a_2\cdots)$ must satisfy Eq.\ (\ref{app_A_a}) for all distributions $\mathbf{n}$ in the vicinity of the MPD. This can only be if $a_i=\text{const}$. It follows $A=a_i=\text{const.}$ 
Now, from Eq.\ (\ref{app_A_def}) we note that the numerical value of $A$ multiplies all $W$ by that factor.  Treating Eq.\ (\ref{app_a_def}) as a recursion for $\ww_i$, we see that the numerical value of $a$ multiplies $\ww_i$ by the factor $a^{i-1}$. Since $A=a$, and Eq.\ (\ref{app_homo}) must be satisfied, we conclude $A=a=1$. 

\subsection{Solution for the Product Kernel}
With $k_{ij}=ij$, the mean kernel in a distribution $\mathbf{n}$ of the $(M,N)$ ensemble is
\begin{equation*}
\kk{n} 
    =   \frac{\sum_i\sum_j ij n_i n_j}{N(N-1)}
    -   \frac{\sum_i i^2 n_i}{N(N-1)}  
    \to \frac{\sum_i\sum_j ij n_i n_j}{N(N-1)} .
\end{equation*}
Asymptotically, 
\begin{equation*}
    \kk{n} \to \ens{k_{M,N}} \to \left(\frac{M}{N}\right)^2 .
\end{equation*}
Applying this result to Eq.\ (\ref{PF}) we obtain
\begin{equation*} 
\label{agg_CL_prodK_PF}
\Omega_{M,N}^\text{prod} = 
   \left(N!\,\frac{M^{M-N}}{M!}\right)^2 
   \binom{M-1}{N-1}.   
\end{equation*}
The parameters $\beta$ and $q$ are
\begin{multline*}
   \beta  = \log\frac{\Omega_{M+1,M}}{\Omega_{M,N}} =  \\ 
          \log \left(\frac{M^{-2(M-N)+1} (M+1)^{2(M-N)}}{M-N+1}\right) \\ 
          \to 1-\frac{N}{M} - 2\log\left(1-\frac{N}{M}\right) ,
\end{multline*}
and
\begin{align*}
    q &= \frac{\Omega_{M,N+1}}{\Omega_{M,N}} 
      =  
       \frac{(N+1)^2 (M-N)}{M^2 N}   \\ 
      &\to 
      \frac{N}{M}\left(1-\frac{N}{M}\right) .
\end{align*}
From Eq.\ (\ref{app_a_def}) with $a_i=1$ and $\kk{n'}/\ens{k_{M,N+1}}=1$ we obtain a recursion for $w_i$:
\begin{equation}
   \ww_i = \frac{1}{i-1}\sum_{j=1}^{i-1} (i-j)j \ww_{i-j} \ww_j ,
\end{equation}
whose inversion gives \citep{Leyvraz:PR03}
\begin{equation}
   \ww_i =2 \frac{(2 i)^{i-2}}{i!} .
\end{equation}

\begin{thebibliography}{28}
\expandafter\ifx\csname natexlab\endcsname\relax\def\natexlab#1{#1}\fi
\expandafter\ifx\csname bibnamefont\endcsname\relax
  \def\bibnamefont#1{#1}\fi
\expandafter\ifx\csname bibfnamefont\endcsname\relax
  \def\bibfnamefont#1{#1}\fi
\expandafter\ifx\csname citenamefont\endcsname\relax
  \def\citenamefont#1{#1}\fi
\expandafter\ifx\csname url\endcsname\relax
  \def\url#1{\texttt{#1}}\fi
\expandafter\ifx\csname urlprefix\endcsname\relax\def\urlprefix{URL }\fi
\providecommand{\bibinfo}[2]{#2}
\providecommand{\eprint}[2][]{\url{#2}}

\bibitem[{\citenamefont{Leyvraz}(2003)}]{Leyvraz:PR03}
\bibinfo{author}{\bibfnamefont{F.}~\bibnamefont{Leyvraz}},
  \bibinfo{journal}{Physics Reports} \textbf{\bibinfo{volume}{383}},
  \bibinfo{pages}{95} (\bibinfo{year}{2003}), \urlprefix\url{http://dx.doi.org
  /10.1016/S0370-1573(03)00241-2}.

\bibitem[{\citenamefont{Stockmayer}(1943)}]{Stockmayer:JCP43}
\bibinfo{author}{\bibfnamefont{W.~H.} \bibnamefont{Stockmayer}},
  \bibinfo{journal}{The Journal of Chemical Physics}
  \textbf{\bibinfo{volume}{11}}, \bibinfo{pages}{45} (\bibinfo{year}{1943}),
  \urlprefix\url{http://link.aip.org/link/?JCP/11/45/1}.

\bibitem[{\citenamefont{Aldous}(1999)}]{Aldous:B99}
\bibinfo{author}{\bibfnamefont{D.~J.} \bibnamefont{Aldous}},
  \bibinfo{journal}{Bernoulli} \textbf{\bibinfo{volume}{5}}, \bibinfo{pages}{3}
  (\bibinfo{year}{1999}).

\bibitem[{\citenamefont{Smoluchowski}(1917)}]{Smoluchowski:ZFPC17}
\bibinfo{author}{\bibfnamefont{M.}~\bibnamefont{Smoluchowski}},
  \bibinfo{journal}{Zeitschrift f{\"u}r Physikalische Chemie}
  \textbf{\bibinfo{volume}{92}}, \bibinfo{pages}{129} (\bibinfo{year}{1917}).

\bibitem[{\citenamefont{Smoluchowski}(1916)}]{Smoluchowski:PZ16}
\bibinfo{author}{\bibfnamefont{M.}~\bibnamefont{Smoluchowski}},
  \bibinfo{journal}{Physik. Zeitschr.} \textbf{\bibinfo{volume}{17}},
  \bibinfo{pages}{585} (\bibinfo{year}{1916}).

\bibitem[{\citenamefont{Friedlander}(2000)}]{Friedlander:00}
\bibinfo{author}{\bibfnamefont{S.~K.} \bibnamefont{Friedlander}},
  \emph{\bibinfo{title}{Smoke, Dust and Haze}} (\bibinfo{publisher}{Oxford
  University Press}, \bibinfo{address}{New York}, \bibinfo{year}{2000}),
  \bibinfo{edition}{2nd} ed.

\bibitem[{\citenamefont{Gueron and Levin}(1995)}]{Gueron:MB95}
\bibinfo{author}{\bibfnamefont{S.}~\bibnamefont{Gueron}} \bibnamefont{and}
  \bibinfo{author}{\bibfnamefont{S.~A.} \bibnamefont{Levin}},
  \bibinfo{journal}{Mathematical Biosciences} \textbf{\bibinfo{volume}{128}},
  \bibinfo{pages}{243} (\bibinfo{year}{1995}),
  \urlprefix\url{http://www.sciencedirect.com/science/article/pii/002555649400074A}.

\bibitem[{\citenamefont{Drake}(1972)}]{Drake:72}
\bibinfo{author}{\bibfnamefont{R.~L.} \bibnamefont{Drake}}, in
  \emph{\bibinfo{booktitle}{Topics in current aerosol research}}, edited by
  \bibinfo{editor}{\bibfnamefont{C.~M.} \bibnamefont{Hidy}} \bibnamefont{and}
  \bibinfo{editor}{\bibfnamefont{J.~R.} \bibnamefont{Brock}}
  (\bibinfo{publisher}{Pergamon Press, New York}, \bibinfo{year}{1972}),
  vol.~\bibinfo{volume}{3}, pp. \bibinfo{pages}{204--379}.

\bibitem[{\citenamefont{Flory}(1941)}]{Flory:JACS41b}
\bibinfo{author}{\bibfnamefont{P.~J.} \bibnamefont{Flory}},
  \bibinfo{journal}{Journal of the American Chemical Society}
  \textbf{\bibinfo{volume}{63}}, \bibinfo{pages}{3091} (\bibinfo{year}{1941}),
  \eprint{http://pubs.acs.org/doi/pdf/10.1021/ja01856a062},
  \urlprefix\url{http://pubs.acs.org/doi/abs/10.1021/ja01856a062}.

\bibitem[{\citenamefont{Erd{\H o}s and R{\'e}nyi}(1960)}]{Erdos:PMIHA60}
\bibinfo{author}{\bibfnamefont{P.}~\bibnamefont{Erd{\H o}s}} \bibnamefont{and}
  \bibinfo{author}{\bibfnamefont{A.}~\bibnamefont{R{\'e}nyi}},
  \bibinfo{journal}{Publ. Math. Inst. Hung. Acad. Sci}
  \textbf{\bibinfo{volume}{5}}, \bibinfo{pages}{17} (\bibinfo{year}{1960}).

\bibitem[{\citenamefont{Krapivsky et~al.}(2010)\citenamefont{Krapivsky, Redner,
  and Ben-Naim}}]{KrapivskyRednerBenNaim}
\bibinfo{author}{\bibfnamefont{P.~L.} \bibnamefont{Krapivsky}},
  \bibinfo{author}{\bibfnamefont{S.}~\bibnamefont{Redner}}, \bibnamefont{and}
  \bibinfo{author}{\bibfnamefont{E.}~\bibnamefont{Ben-Naim}},
  \emph{\bibinfo{title}{A Kinetic View of Statistical Physics}}
  (\bibinfo{publisher}{Cambridge University Press},
  \bibinfo{address}{Cambridge}, \bibinfo{year}{2010}).

\bibitem[{\citenamefont{Ziff and Stell}(1980)}]{Ziff:JCP80}
\bibinfo{author}{\bibfnamefont{R.~M.} \bibnamefont{Ziff}} \bibnamefont{and}
  \bibinfo{author}{\bibfnamefont{G.}~\bibnamefont{Stell}},
  \bibinfo{journal}{The Journal of Chemical Physics}
  \textbf{\bibinfo{volume}{73}}, \bibinfo{pages}{3492} (\bibinfo{year}{1980}),
  \urlprefix\url{http://link.aip.org/link/?JCP/73/3492/1}.

\bibitem[{\citenamefont{Matsoukas}(2014{\natexlab{a}})}]{Matsoukas:statPBE_PRE14}
\bibinfo{author}{\bibfnamefont{T.}~\bibnamefont{Matsoukas}},
  \bibinfo{journal}{Phys. Rev. E} \textbf{\bibinfo{volume}{90}},
  \bibinfo{pages}{022113} (\bibinfo{year}{2014}{\natexlab{a}}),
  \urlprefix\url{http://link.aps.org/doi/10.1103/PhysRevE.90.022113}.

\bibitem[{\citenamefont{Marcus}(1968)}]{Marcus:T68}
\bibinfo{author}{\bibfnamefont{A.}~\bibnamefont{Marcus}},
  \bibinfo{journal}{Technometrics} \textbf{\bibinfo{volume}{10}},
  \bibinfo{pages}{133} (\bibinfo{year}{1968}),
  \urlprefix\url{http://www.jstor.org/stable/1266230}.

\bibitem[{\citenamefont{Lushnikov}(1978)}]{Lushnikov:JCIS78}
\bibinfo{author}{\bibfnamefont{A.~A.} \bibnamefont{Lushnikov}},
  \bibinfo{journal}{Journal of Colloid and Interface Science}
  \textbf{\bibinfo{volume}{65}}, \bibinfo{pages}{276} (\bibinfo{year}{1978}),
  \urlprefix\url{dx.doi.org/10.1016/0021-9797(78)90158-3}.

\bibitem[{\citenamefont{Lushnikov}(2004)}]{Lushnikov:PRL04}
\bibinfo{author}{\bibfnamefont{A.~A.} \bibnamefont{Lushnikov}},
  \bibinfo{journal}{Phys. Rev. Lett.} \textbf{\bibinfo{volume}{93}},
  \bibinfo{pages}{198302} (\bibinfo{year}{2004}),
  \urlprefix\url{http://link.aps.org/doi/10.1103/PhysRevLett.93.198302}.

\bibitem[{\citenamefont{Lushnikov}(2005)}]{Lushnikov:PRE05}
\bibinfo{author}{\bibfnamefont{A.~A.} \bibnamefont{Lushnikov}},
  \bibinfo{journal}{Phys. Rev. E} \textbf{\bibinfo{volume}{71}},
  \bibinfo{pages}{046129} (\bibinfo{year}{2005}).

\bibitem[{\citenamefont{Spouge}(1983{\natexlab{a}})}]{Spouge:M83_831}
\bibinfo{author}{\bibfnamefont{J.~L.} \bibnamefont{Spouge}},
  \bibinfo{journal}{Macromolecules} \textbf{\bibinfo{volume}{16}},
  \bibinfo{pages}{831} (\bibinfo{year}{1983}{\natexlab{a}}),
  \urlprefix\url{http://pubs.acs.org/doi/abs/10.1021/ma00239a021}.

\bibitem[{\citenamefont{Hendriks et~al.}(1985)\citenamefont{Hendriks, Spouge,
  Eibl, and Schreckenberg}}]{Hendriks:ZPCM85B}
\bibinfo{author}{\bibfnamefont{E.~M.} \bibnamefont{Hendriks}},
  \bibinfo{author}{\bibfnamefont{J.~L.} \bibnamefont{Spouge}},
  \bibinfo{author}{\bibfnamefont{M.}~\bibnamefont{Eibl}}, \bibnamefont{and}
  \bibinfo{author}{\bibfnamefont{M.}~\bibnamefont{Schreckenberg}},
  \bibinfo{journal}{Zeitschrift f{\"u}r Physik B Condensed Matter}
  \textbf{\bibinfo{volume}{58}}, \bibinfo{pages}{219} (\bibinfo{year}{1985}),
  \urlprefix\url{http://dx.doi.org/10.1007/BF01309254}.

\bibitem[{\citenamefont{Durrett et~al.}(1999)\citenamefont{Durrett, Granovsky,
  and Gueron}}]{Durrett:JTP99}
\bibinfo{author}{\bibfnamefont{R.}~\bibnamefont{Durrett}},
  \bibinfo{author}{\bibfnamefont{B.~L.} \bibnamefont{Granovsky}},
  \bibnamefont{and} \bibinfo{author}{\bibfnamefont{S.}~\bibnamefont{Gueron}},
  \bibinfo{journal}{Journal of Theoretical Probability}
  \textbf{\bibinfo{volume}{12}}, \bibinfo{pages}{447} (\bibinfo{year}{1999}),
  ISSN \bibinfo{issn}{0894-9840}, \bibinfo{note}{10.1023/A:1021682212351},
  \urlprefix\url{http://dx.doi.org/10.1023/A:1021682212351}.

\bibitem[{\citenamefont{Berestycki and Pitman}(2007)}]{Berestycki:07}
\bibinfo{author}{\bibfnamefont{N.}~\bibnamefont{Berestycki}} \bibnamefont{and}
  \bibinfo{author}{\bibfnamefont{J.}~\bibnamefont{Pitman}},
  \bibinfo{journal}{Journal of Statistical Physics}
  \textbf{\bibinfo{volume}{127}}, \bibinfo{pages}{381} (\bibinfo{year}{2007}),
  \urlprefix\url{http://dx.doi.org/10.1007/s10955-006-9261-1}.

\bibitem[{\citenamefont{Duerr and Dietz}(2000)}]{Duerr:MB00}
\bibinfo{author}{\bibfnamefont{H.~P.} \bibnamefont{Duerr}} \bibnamefont{and}
  \bibinfo{author}{\bibfnamefont{K.}~\bibnamefont{Dietz}},
  \bibinfo{journal}{Mathematical Biosciences} \textbf{\bibinfo{volume}{165}},
  \bibinfo{pages}{135} (\bibinfo{year}{2000}).

\bibitem[{\citenamefont{Sheth and Pitman}(1997)}]{Sheth:MNRAS97}
\bibinfo{author}{\bibfnamefont{R.~K.} \bibnamefont{Sheth}} \bibnamefont{and}
  \bibinfo{author}{\bibfnamefont{J.}~\bibnamefont{Pitman}},
  \bibinfo{journal}{Mon. Not. R. Astron. Soc.} \textbf{\bibinfo{volume}{289}},
  \bibinfo{pages}{66} (\bibinfo{year}{1997}).

\bibitem[{\citenamefont{Matsoukas}(2014{\natexlab{b}})}]{Matsoukas:statPBE2}
\bibinfo{author}{\bibfnamefont{T.}~\bibnamefont{Matsoukas}},
  \bibinfo{journal}{ArXiv} p. \bibinfo{pages}{(in preparation)}
  (\bibinfo{year}{2014}{\natexlab{b}}).

\bibitem[{\citenamefont{Pitman}(2006)}]{Pitman:06}
\bibinfo{author}{\bibfnamefont{J.}~\bibnamefont{Pitman}},
  \emph{\bibinfo{title}{Combinatorial Stochastic Processes}}, vol.
  \bibinfo{volume}{1875} (\bibinfo{publisher}{Springer}, \bibinfo{year}{2006}).

\bibitem[{\citenamefont{Smith and Matsoukas}(1998)}]{Matsoukas:CES98}
\bibinfo{author}{\bibfnamefont{M.}~\bibnamefont{Smith}} \bibnamefont{and}
  \bibinfo{author}{\bibfnamefont{T.}~\bibnamefont{Matsoukas}},
  \bibinfo{journal}{Chemical Engineering Science}
  \textbf{\bibinfo{volume}{53}}, \bibinfo{pages}{1777 } (\bibinfo{year}{1998}),
  ISSN \bibinfo{issn}{0009-2509},
  \urlprefix\url{http://www.sciencedirect.com/science/article/pii/S0009250998000451}.

\bibitem[{\citenamefont{Spouge}(1985)}]{Spouge:JPMG85}
\bibinfo{author}{\bibfnamefont{J.~L.} \bibnamefont{Spouge}},
  \bibinfo{journal}{Journal of Physics A: Mathematical and General}
  \textbf{\bibinfo{volume}{18}}, \bibinfo{pages}{3063} (\bibinfo{year}{1985}),
  \urlprefix\url{http://stacks.iop.org/0305-4470/18/i=15/a=028}.

\bibitem[{\citenamefont{Spouge}(1983{\natexlab{b}})}]{Spouge:M83_121}
\bibinfo{author}{\bibfnamefont{J.~L.} \bibnamefont{Spouge}},
  \bibinfo{journal}{Macromolecules} \textbf{\bibinfo{volume}{16}},
  \bibinfo{pages}{121} (\bibinfo{year}{1983}{\natexlab{b}}),
  \urlprefix\url{http://pubs.acs.org/doi/abs/10.1021/ma00235a024}.

\end{thebibliography}

\end{document}